\documentstyle[aps,eqsecnum,epsfig,rotating]{revtex}

\title{\hfill \begin{small} Preprint: WSU--NP--96--16 \end{small}\\
A theoretical view of practical problems in interferometry}

\author{A.~Makhlin, E.~Surdutovich, and G.~Welke\footnote{Submitted by
    G.~Welke for Proceedings, HIPAGS '96 Workshop, August 22--24, 1996, 
    Wayne State University, Detroit, Michigan}}
\address{Department of Physics and Astronomy, Wayne State
University, Detroit MI 48202, USA}

\begin{document}

\maketitle

\begin{abstract}
  Interferometry is discussed in terms of the representation of the
  source. In particular, scale--invariant 1--d hydrodynamics is
  revisited, and extended to the case of unequal transverse masses. It
  is argued that that kaon emission occurs over a short time interval.
  Exact results for models of two-- and three--dimensional flow are
  presented, which exhibit altered scaling laws.  Such qualitative
  trends, together with other observables, are vital if one is to draw
  conclusions about the source [\ref{welke:long}].
\end{abstract}

\section{Introduction}\label{welke_sec:intro}

The discovery a quark--gluon plasma will to some extent be linked to
the measurement of the geometric size and other macroscopic
characteristics of the reaction zone. An important tool in
accomplishing this is interferometry. As proposed by Hanbury-Brown and
Twiss to measure stellar sizes, and as the GGLP effect in (low energy)
nuclear physics, it provides for a straightforward determination of
the size of the ``hot source.'' In high--energy collisions,
on the other hand, the correspondence between measured quantities and
parameters of the emitting system is less clear [\ref{welke:review}].
For example, correlations on the same scale as the size of the
emitting system make an interferometric study of the correlations
inside the matter and a measurement of the external geometry
exceedingly difficult [\ref{welke:Mak72}].

For A--A collisions, even in the simplest scenario of hydrodynamic
evolution, the relation between the inclusive one-- and two--particle
spectra and the parameters of the emitting system does not follow the
classical scheme of interferometry [\ref{welke:Scott}]. Since no
direct inverse solution exists, and the question of {\em what} is
measured by the two--pion correlation function becomes nontrivial.
The problem for interferometry is thus to determine the parameters of
a judiciously chosen density matrix $\rho$ for the system, by
measuring the inclusive cross-sections $dN_{1}/d{\vec k}$,
$dN_{2}/d{\vec k}_{1} d{\vec k}_{2},\ldots$~.  Here, we shall restrict
ourselves to particle sources that may be described in terms of
(semi--) classical one--particle distributions, or even by only a few
macroscopic parameters.  Practically, current dynamical models produce
precisely such sources. However, even with these restrictions, the
two--particle spectra are not easily decoded. We therefore suggest in
Sect~\ref{welke_sec:1hydro} an illustrative ``step-by-step'' strategy:
Essentially, the data are first confronted with the simplest
physically motivated model, and discrepancies are then used show
directions for improvement.  Ideally, this requires that the model
should have an ``adequate'' (semi--) analytical solution for the
correlator, allowing one to obtain the model parameters precisely and
directly, without resorting to intermediate fits to the data.

Recent SPS experiments [\ref{welke:NA35},\ref{welke:NA44}] have
observed $m_\perp$ scaling of longitudinal radii, as well as the
$y$--dependence of the correlator width predicted by boost-invariant
1--d hydrodynamic expansion [\ref{welke:MS}]. This provides a strong
reason for taking macroscopic collective behavior ({\it i.e.},
hydrodynamics) seriously and using interferometry to study it -- at
least at SPS energies.  Here, we shall therefore extend the
calculation of Ref.~[\ref{welke:MS}] to give the (analytic) solution
in the case of unequal transverse momenta, and illustrate why the data
support a relatively rapid freeze--out picture for both pions and
kaons.  We also discuss natural extensions of the model to one that
might be more appropriate at the AGS: (1) 1--d Landau hydrodynamics,
and (2) some exactly calculable examples with two-- and
three--dimensional flow. In these cases, the scaling law is altered,
and we argue generally that such qualitative trends, together with
other observables, are important in establishing an adequate model of
of the source, and only then the value of its parameters.

Before turning our attention to these matters, however, we discuss a
related issue: there is no full consensus on what mathematical
formulation is adequate for applications of interferometry.  For
example, an analysis of polarization effects in photon interference
[\ref{welke:Phot}] indicates that the languages of locally defined
states and Wigner phase-space distributions lead to different answers;
there is a practical need to resolve such controversies.  We shall
therefore begin in Sec.~\ref{welke_sec:SN2} by briefly formulating the
problem of interferometry for an expanding source at the {\em
  operator} level.  This emphasizes the quantum nature of the problem,
and allows us to solve (correctly) the basic problem of one-- and
two--particle propagation without having to address the unknown nature
of the source. With these expressions in hand, we discuss three
examples of source input.

\section{Preliminaries}\label{welke_sec:SN2}

Let $|{\rm in}\rangle$ be any one of the initial states of a system
emitting a pion field ${\hat \varphi}(x)$ at time $t_c$.  This field
is detected later by an analyzer, tuned to the measurement of the
momentum of the free pion; its eigenfunctions are thus plane waves
$f_{\vec k}(x)$ with $x^0>t_c$, and the corresponding annihilation
operators $A_{\vec k}$ for momentum ${\vec k}$ in the final state are
given by 
\begin{eqnarray}
{\hat A}_{\vec k}\;=\;\int d^{3}x\; f^{*}_{\vec k}(x)\,i\!
\stackrel{\leftrightarrow}{\partial}_{x}^{\,0}
{\hat \varphi}(x)~ ~ ~, \label{welke_eq:E2.2}
\end{eqnarray}
\noindent where $(a\!\!\stackrel{\leftrightarrow}{\partial}_{x}^{\,0}
\!\!b)\equiv a(\partial_x^{\,0}\,b) - (\partial_x^{\,0}\,a)b$.  The
operator ${\hat A}_{\vec k}$ describes the effect of a detector far
from the point of emission, so, by definition, the pion is detected
on-mass-shell.  The inclusive amplitudes to find one pion with momentum
${\vec k}$, or two with momenta ${\vec k}_1$, ${\vec k}_2$ in the
final state are $\langle X|\,{\hat A}_{\vec k}{\hat S}\,|{\rm in}
\rangle$ and $\langle X|\,{\hat A}_{{\vec k}_2}
{\hat A}_{{\vec k}_1}\;{\hat S}\,|{\rm in}\rangle$,
respectively. Here, ${\hat S}$ is the evolution operator after
freeze--out, and the $|X\rangle$ form a complete set of all possible
secondaries. One obtains the one-- and two--particle inclusive spectra
by summing the squared modulus of these amplitudes over all
(undetected) states $|X\rangle$, and averaging over the initial
ensemble $\rho$.

An analyzer tuned to ${\vec k}$ for singles, or  ${\vec k}_1$, ${\vec
  k}_2$ for pairs, is represented by the operators
\begin{eqnarray}
{\hat N}(\vec k)\; =\; {\hat A}^\dagger_{\vec
k}{\hat A}_{\vec k}~,~ ~ ~ ~{\rm and}~ ~ ~ ~
{\hat N}({\vec k}_1,{\vec k}_2)\;=\;
{\hat A}^\dagger_{{\vec k}_{1}} {\hat A}^\dagger_{{\vec k}_{2}}
{\hat A}_{{\vec k}_{2}} {\hat A}_{{\vec k}_{1}}\;=\;
{\hat N}({\vec k}_1) ({\hat N}({\vec k}_2)-
\delta({\vec k}_1-{\vec k}_2) )~,\label{welke_eq:E2.8}
\label{welke_eq:E2.03}
\end{eqnarray}
respectively. In the simplest description,
the pion field ${\hat \varphi}(x)$ reaches the detector after free
propagation
\begin{eqnarray}
{\hat \varphi}(x) \;=\; \int d\Sigma_{\mu}(y)\:G_{\rm ret}(x-y)
\stackrel{\leftrightarrow}{\partial}^{\mu}_{y}
{\hat \varphi}(y)~ ~ ~,\label{welke_eq:E2.9}
\end{eqnarray}
\noindent where the field ${\hat
  \varphi}(y)$ and the space of states in which the density matrix
acts are defined on the freeze--out hypersurface $\Sigma_c$.
The pion Fock operator (\ref{welke_eq:E2.2}) may now be expressed in
terms of the initial fields, and simplified using the explicit form of
the free pion propagators:
\begin{eqnarray}
{\hat S}^\dagger {\hat A}_{\vec k}\,{\hat S}
&=& \theta(x^0-y^0)\int d\Sigma_{\mu}(y)
\; f^*_{\vec k}(y) i\!\stackrel{\leftrightarrow}{\partial}^{\mu}_{y}
{\hat \varphi} (y)~ ~ ~.\label{welke_eq:E2.12}
\end{eqnarray}
\noindent Clearly, the analyzers perform an on--shell Fourier
expansion of the initial data.  Collecting these results, we may
rewrite the number operators for single and pairs as
\begin{eqnarray}
{\hat N}({\vec k}) &=& \theta(x^0\!-\!y^0)\int\! d\Sigma_{\mu}(y_1)\,
d\Sigma_{\nu}(y_2)\:\bigg [ f_{\vec
k}(y_1)\,i\!\stackrel{\leftrightarrow}{\partial}^{\,\mu}_{y_1}
{\hat \varphi}^\dagger
(y_1)\bigg ]\:\bigg [ {\hat \varphi} (y_2)\, i\!
\stackrel{\leftrightarrow}{\partial}^{\,\nu}_{y_2} 
f^{*}_{\vec k}(y_2)\bigg ]~ ~ ~\label{welke_eq:E2.13} \\
{\hat N}({\vec k}_1,{\vec k}_2)
&=& \theta(x^0-y^0)\int d\Sigma_{\mu}(y_1)\,d\Sigma_{\nu}(y_3)\,
d\Sigma_{\rho}(y_4)\,d\Sigma_{\lambda}(y_2)  \nonumber\\
&\times&\bigg [f_{{\vec k}_1}(y_1)f_{{\vec k}_2}(y_3) 
\stackrel{\leftrightarrow}{\partial}^{\,\mu}_{y_1}
\stackrel{\leftrightarrow}{\partial}^{\,\nu}_{y_3}
{\hat \varphi}^\dagger (y_1){\hat \varphi}^\dagger (y_3) \bigg ]\:
\bigg [ {\hat \varphi}
(y_4){\hat \varphi}
(y_2)\stackrel{\leftrightarrow}{\partial}^{\,\rho}_{y_4}
\stackrel{\leftrightarrow}{\partial}^{\,\lambda}_{y_2}
f^{*}_{{\vec k}_2}(y_4) f^{*}_{{\vec k}_1}(y_2)\bigg ]~,
\label{welke_eq:E2.14}
\end{eqnarray}
respectively. The field operator products in these
equations are now to be averaged over the density
matrix of initial states. The interferometry problem
requires that the basis the final {\em quantum}
states of the preceding stage of evolution should be described
explicitly. We give three examples: 

\noindent{\bf 1. Emission from two one--dimensional 
  cavities:} \label{welke_subsec:TM1} We have two
sets of the eigenstates
\begin{eqnarray}
\phi_{ p,N}(y) &=& (2a)^{-1/2}(2p_{0})^{-1/2}
e^{-i\,p\cdot y}~,~ ~ ~ ~{\rm for}~L_N-a\le y \le L_N+a~, ~ ~{\rm where~}
L_N=\pm L \nonumber \\ &=& 0~,~ ~ ~ ~{\rm otherwise}. \label{welke_eq:E4.13}
\end{eqnarray}
The field decomposition ${\hat \varphi}(y)\;=\;\sum_{p,N}
{\hat a}_{p,N}\:\phi_{p,N}(y)$
acquires an additional index enumerating
the cavities, and states belonging to different cavities are independent:
$[{\hat a}_{p,N},\, {\hat a}^\dagger_{p^\prime,N^\prime} ]\;=\;
\delta_{pp^\prime}\delta_{NN^\prime}$. From Eq.~(\ref{welke_eq:E2.13}),
the one--particle distribution for when the walls are
removed at $y_0=0$ is obtained as
\begin{eqnarray}
\langle {\hat N}(k)\rangle \;=\;\sum_N \sum_{p} \: n(p,N)  \:
{(\omega_k + \omega_p)^2 \over 4\omega_k \omega_p }\;
{\sin^2 (p-k)a \over \pi a (p-k)^2  }~ ~ ~,\label{welke_eq:E3.7}
\end{eqnarray}
where $n(p,N)= \langle {\hat a}^\dagger_{p,N}{\hat a}_{p,N}\rangle$ 
is the boson occupation number in cavity $N$.

For the two-particle spectrum there is interference between four
amplitudes: two when each particle is emitted from a different cavity,
and two more when both particles originate from the same cavity.  For
the statistical average of Eq.~(\ref{welke_eq:E2.14}) we choose
\begin{eqnarray}
  \langle {\hat a}^\dagger_1 {\hat a}^\dagger_3 {\hat a}_4 {\hat a}_2
  \rangle=n(1) n(3)\:\bigg [\delta_{p_1p_2}\delta_{N_1N_2}\delta_{p_3
    p_4} \delta_{N_3 N_4} + \delta_{p_1p_4}\delta_{N_1N_4}\delta_{p_3
    p_2}\delta_{N_3 N_2}\bigg]~, \label{welke_eq:E3.18}
\end{eqnarray} 
where we have introduced the notation $i \equiv (p_i,N_i)$. Then
\begin{eqnarray}
\lefteqn{\langle {\hat N}(k_1,k_2)\rangle = \langle {\hat
      N}(k_1)\rangle\: \langle {\hat N}(k_2)\rangle + \sum_{1,3}\;
    {(\omega_{k_1}\!+\!\omega_{p_1})(\omega_{k_1}\!+\!\omega_{p_3})
      (\omega_{k_2}\!+\!\omega_{p_1})(\omega_{k_2}\!+\!\omega_{p_3})
      \over 4\pi^2 4a^2 4 \omega_{k_1} \omega_{k_2}
      \omega_{p_1}\omega_{p_3}}\;\times} \label{welke_eq:E3.19}\\ 
&& n(1) n(3) {\sin (p_1-k_1)a \over (p_1-k_1)}\,{\sin (p_1-k_2)a \over
    (p_1-k_2)} \,{\sin (p_3-k_1)a \over (p_3-k_1)}\,{\sin (p_3-k_2)a
    \over (p_3-k_2)} {\rm e}^{-i(L_1-L_3)(k_1-k_2)}~,\nonumber
\end{eqnarray} 
where $L_j=L(N_j)=\pm L$, and momentum $p_j$ originates from the
cavity $N_j$.  The two terms in the sum with $N_1\neq N_3$ lead to the
usual interference term, with $\Delta k \sim
1/(2L)$.  The average of the product of Fock operators
(\ref{welke_eq:E3.18}), {\em i.e.}, correlations in the emitting
system, depends only on ``internal'' variables which identify the
stationary states in the cavities before they are opened.  On the
other hand, the $\sin$--functions serve as the projectors of these
states onto the states of free propagation in which the particles are
detected, and require additional physical input. First, detector
resolution should not allow one to determine which cavity the particle
originated from, {\it i.e.}, we do not have $|k_1-k_2|\:L\; \gg\; 1$.
This condition corresponds to the Rayleigh criterium: we
are not able to construct an ``optical'' image of source. Secondly, in
most physical situations $L\gg a$, so that $|k_1-k_2|\:a\; \ll\; 1$.
This inequality removes the momentum integral in
Eq.~(\ref{welke_eq:E3.7}), so that a measurement of the one-particle
spectrum does not allow for a determination of the cavity size.

\noindent {\bf 2. Emission from many cavities and hydrodynamics:}
To generalize the two--cavity model to the freeze--out of an extended
hydrodynamic system, we consider a continuous set of decaying thermal
cells, defined on a freeze--out surface $T(x)=T_c$ and Doppler shifted
to account for cell motion [\ref{welke:MS}].  As in the previous
example, the initial quantum states in one cell are assumed to commute
with states in a different cell.  This additional information implies
the existence of two length scales, and allows us to use a
quasi--classical picture.  Within a cell, the system is governed by
the short-range dynamics conveniently described in momentum language
and if the interactions are switched off, the particles begin to
propagate freely with the frozen momentum distribution defined {\it
  within the cell}. To the lowest approximation of the dynamical
interactions and in agreement with the freeze-out concept, these
particles are on-mass-shell.  In terms of an auxiliary ``emission
function'' $J(k_1,k_2)$,
\begin{eqnarray}
  J(k_1,k_2)\;\equiv\;\int_{\Sigma_{c}} d\Sigma_{\mu}(x)\: {k^{\mu}_{1}+
    k^{\mu}_{2}\over{2}}\: n(k_{1}\cdot u(x))\: {\rm
    e}^{-i(k_{1}-k_{2})x}~ ~ ~,\label{welke_eq:E3.24}
\end{eqnarray}              
one obtains for the one-- and two--particle spectra [\ref{welke:MS}]:
\begin{eqnarray}
k^{0}{{dN_{1}} \over {d{\vec k}} } = J(k,k)~,~{\rm and}~ ~ ~ ~
k^{0}_{1} k^{0}_{2}{{dN_{2}} \over {d{\vec k}_{1} d{\vec k}_{2}}}
= J(k_{1},k_{1}) J(k_{2},k_{2})+ {\rm Re}\bigg
[J(k_{1},k_{2})J(k_{2},k_{1})\bigg ]~,
\label{welke_eq:E3.26}
\end{eqnarray}
%\label{welke_eq:E3.25} \\
respectively. The latter equation is explicitly
symmetric in $x_1$ and $x_2$,  as well as in $k_1$ and $k_2$.
It reveals interference of the two amplitudes for the two--pion state
(consisting of the two on-mass-shell pions with quantum numbers ${\vec
  k}_1$ and ${\vec k}_2$) to be emitted by the two sources. These
sources are located at points $x_1$ and $x_2$ and have true thermal
distributions, $n(k_i\cdot u(x_i))$, with
on-shell particles.

\noindent {\bf 3. Wigner formalism:}
Simulations of heavy ion collisions typically produce a final state in
terms of Wigner distributions, so it is of practical value
to examine the two--particle spectrum in terms of them.  Instead of
Eq.~(\ref{welke_eq:E3.18}), we may express the independence of the
dynamical processes in the emitting system via
\begin{eqnarray}
  \langle {\hat \varphi}^{\dag}(x){\hat \varphi}^{\dag}(x'){\hat
    \varphi}(y){\hat \varphi}(y')\rangle = \langle {\hat
    \varphi}^{\dag}(x){\hat \varphi}(y)\rangle \langle {\hat
    \varphi}^{\dag}(x'){\hat \varphi}(y')\rangle + \langle {\hat
    \varphi}^{\dag}(x){\hat \varphi}(y')\rangle \langle {\hat
    \varphi}^{\dag}(x'){\hat \varphi}(y)\rangle
\label{welke_eq:E3.23a}
\end{eqnarray}
\noindent which implies that the distance $z=x-y$ between the points in one
correlator does not exceed a correlation length $a_{cor}$ which, in
turn, is much less than the full size $L$ of the system. In the first
term, for example, $x,y\sim R_1,~x-y<a_{cor}$ and $x',y'\sim
R_2,~x'-y'<a_{cor}$, while $R_1-R_2\gg a_{cor}$.  Under these
conditions, it is possible to think of the coordinates $R_j$ as labels
of domains of non-vanishing correlation, while the Fourier transform
over ${\vec z}$ yields the local spectrum. Thus, changing
variables\footnote{A formal change of variables can already be made at
  the operator level, Eq.~(\protect\ref{welke_eq:E2.14}), but the
  calculation of the two-particle spectrum then becomes ambiguous
  [\protect\ref{welke:long}].~ ~ $^2$The problem is even more acute
  for photon emission: unphysical longitudinally polarized states are
  introduced [\protect\ref{welke:long},\protect\ref{welke:Phot}].}
$x=R+z/2,~y=R-z/2$, and introducing the Wigner function ${\cal N}$
\begin{eqnarray}
  \langle {\hat \varphi}^{\dag}(R+{z\over 2}){\hat \varphi}(R-{z\over
    2})\rangle =\int d^3s\: {\cal N}({\vec R},{\vec s}\,)\: {\rm
    e}^{i{\vec s}\cdot {\vec z}}~ ~,
\label{welke_eq:E3.23b}
\end{eqnarray}
\noindent Eqs.~(\ref{welke_eq:E2.14}), (\ref{welke_eq:E3.23a}), and 
(\ref{welke_eq:E3.23b}) yield
\begin{eqnarray}
\lefteqn{\langle {\hat N}(k_1,k_2)\rangle = \int dR_1dz_1ds_1 {\cal N}({\vec
    R}_1,{\vec s}_1) e^{i(s_1-k_1)z_1} \int dR_2dz_2ds_2 {\cal
    N}({\vec R}_2,{\vec s}_2) e^{i(s_2-k_2)z_2} + 
    \label{welke_eq:E3.23c}}\\ && \int\!\!
  dR_1\:dR_2\:dz_1\:dz_2\:ds_1\:ds_2\: {\cal N}({\vec R}_1,{\vec s}_1)
  {\cal N}({\vec R}_2,{\vec s}_2) e^{-i(k_1-k_2)(R_1-R_2)} \:
  {\rm e}^{-i({k_1+k_2\over2}-s_1)z_1-i({k_1+k_2\over2}-s_2)z_2}~.
\nonumber
\end{eqnarray}
\noindent If the range of integration $a_{cor}$ (which, strictly speaking,
depends on $R$) for the $z$--variable is sufficiently large compared
to $1/k_1$, $1/k_2$, the integrals over $z_1$ and $z_2$ approximate to
delta-functions, and the common expression for the correlator $C_2$ in
terms of functions ${\cal N}((k_1+k_2)/2,{\vec R})$ results
[\ref{welke:Scott}]. Physically, however, since the ${\hat
  \varphi}(x)$ are free, there are no states with $s^2\neq m^2$. We
thus face the problem of understanding the origin of a thermal
distribution of off--shell pions, instantly frozen, and detected later
in physical on--shell states.\footnotemark ~ The variables of
integration do not correctly establish the correspondence between
initial and final states.  Moreover, the replacement ${\cal
  N}((k_1+k_2)/2,{\vec R})$ undermines the concept of independent
emission of the two sources -- a necessary condition for interference.
Each source ``knows'' about the momenta of both pions, and one may
reasonably suspect that emission from spatially separated volumes is
not independent.

\section{Interferometry and source models}\label{welke_sec:1hydro}

\noindent {\bf 1. One--dimensional flow:} We revisit interferometry 
for scale invariant 1-d hydrodynamic motion; evidence for such sources
at SPS [\ref{welke:NA35},\ref{welke:NA44}] comes from agreement with
the theoretically predicted [\ref{welke:MS}] $m_t$-- and
$y$--dependence of the ``visible longitudinal size:'' 
\begin{eqnarray}
  R_L \;=\;{\tau \over {\rm cosh}y}\: \sqrt{{T_c \over m_{t}}}~ ~ ~.
\label{welke_eq:E1}
\end{eqnarray}   
\noindent where $y$ is the pair mid-rapidity, and $m_t^2=m^2+p_t^2$.  
Physically, the correlator measures the effective size of the fluid
slice which forms the observable spectrum at a given rapidity.
The larger the transverse momentum of a particle in
the fluid, the more that particle is frozen into collective
longitudinal motion: the subset of particles with high $p_\perp$ will
have a very narrow distribution over $k_{\scriptscriptstyle\|}$.

The dependence (\ref{welke_eq:E1}) was derived for the case of equal
transverse momenta.  This condition is impractical and not the way the
data were taken, so we extend the calculations to the case of unequal
transverse masses. The parameters of the model are the critical
temperature, $T_{c}~ (\sim m_{\pi})$, and the space--like freeze--out
hypersurface, $t^{2}-x^{2}_{\scriptscriptstyle \|} = \tau^{2} = {\rm
  const}$.  We assume a Gaussian transverse distribution
$\exp({-r^{2}_{\bot}/R^{2}_{\bot}})$ of hot matter in a pipe with
effective area $S_{\bot} = \pi R^{2}_{\bot}$.  The particles are
described by their momenta $k^{\mu}_{i} =
(k^{0}_{i},k^{\scriptscriptstyle \|}_{i},{\vec p}_{i}) \equiv
(m_{i}\cosh\theta_{i},\ m_{i}\sinh\theta_{i},\ {\vec p}_{i})$, where
${\vec p}_{i}$ is the transverse momentum, $\theta_{i}$ the particle
rapidity in the beam direction, and $m^2_i = m^{2}_{\pi}+ {\vec
  p}^{\,2}_i$ is the transverse mass. Further, let $2\alpha = \theta_1
- \theta_2$, $2\theta = \theta_{1} + \theta_{2}$ and ${\vec q}_{\bot}
= {\vec p}_{1} - {\vec p}_{2}$.  Approximating the one--particle
distribution by a Boltzmann form, the saddle point method yields from
Eq.~(\ref{welke_eq:E3.24}) (see Ref.~[\ref{welke:long}]):
\begin{eqnarray}
  {{dN}\over{d\theta_1 d{\vec p}_1}} \approx \tau S_{\bot}m_1 \:
  \sqrt{2\pi T_{c}\over{m_{1}}} \: {\rm e}^{-m_{1}/T_c}~ ~ ~,
\label{welke_eq:E4.3}
\end{eqnarray}
\noindent and
\begin{eqnarray}
  \lefteqn{C_2(k_{1},k_{2})-1 = {dN_2\over d\theta_1 d{\vec p}_1 d\theta_2
      d{\vec p}_2}/ \bigg [ {dN_1\over d\theta_1 d{\vec p}_1}
    {{dN_1}\over{d\theta_2 d{\vec p}_2}}\bigg ]
    -1\label{welke_eq:E4.5}}\\ && \approx \frac 14 \; f(|{\vec
    q}_{\bot}|R_{\bot}) \; {{g(z)\,g(1/z)} \over
    {[h(z)\,h(1/z)]^{3/2}}} \exp \bigg \{ -{\mu \over {T_{c}}}\: \bigg
  [ h(z)\cos {H(z)\over{2}} +h({1\over{z}}) \cos
  {{H(1/z)}\over{2}}-z-{1\over z}\bigg ]\bigg \} \nonumber \\ &&
  \times \cos\bigg \{ {\mu \over {T_{c}}}\bigg [h(z)\sin
  {H(z)\over{2}} +h({1\over{z}}) \sin {{H(1/z)}\over{2}} \bigg ] +
  {3\over 4}\bigg [H(z)+H({1\over {z}})\bigg ] +G(z)+G({1\over{z}})
  \bigg \}~,\nonumber
\end{eqnarray} 
\noindent where $\mu =(m_{1}m_{2})^{1/2}$ and $z = (m_{1}/m_{2})^{1/2} $, 
and we have introduced the functions 
\begin{eqnarray}    
  h(z) &\equiv& \bigg \{ \bigg [
  z^{2}-F^{2}(z-{1\over{z}})^{2}+4F^2\sinh^{2}\alpha \bigg ]^2
  +4F^2(z^2-\cosh 2\alpha)^2 \bigg \}^{1/4};~~\nonumber\\ g(z) &\equiv& \bigg
  [(z^{2}+\cosh 2\alpha)^{2} +F^{2}(z^{2}-{1\over{z^{2}}})^{2}\bigg
  ]^{1/2};~~~~~ f(|{\vec q}_{\bot}|R_{\bot})=
{\rm e}^{-{\vec q}^{\,2}_{\bot}R^{2}_{\bot}/2} \\ 
\nonumber\\ \tan H(z) &=& {2F(\cosh 2\alpha-z^{2}) \over
    {z^{2}-F^{2}(z-{1\over {z}})^{2}+4F^{2}\sinh^{2}\alpha}};~~~~~
  \tan G(z) = {F(z^{2}- 1/z^{2}) \over {z^2+\cosh 2\alpha}}~ ~ ~.
\label{welke_eq:E4.6} 
\end{eqnarray}
\noindent with $F=\tau T_c$.
Equation (\ref{welke_eq:E4.5}) is free of the limitation $m_1=m_2$,
and should be used directly to fit the model parameters, $T_c$ and
$\tau$, without an intermediate Gaussian, and at various rapidities.
The $T_c$ may then in turn be confronted with the slope of the
one-particle spectrum.  If, for instance, we find a dependence
$\tau(y)$ then it is natural to consider the effect of a finite
initial size via the 1--d Landau model as a next step.  The visible
size will be given by Eq.~(\ref{welke_eq:E1}), with $\tau \rightarrow
\tau(y)$ [\ref{welke:MS}].  Now, the relevant parameters of the model
are the initial size and temperature, and the shape of the freeze--out
surface.  The general idea is that since there is no unique
prescription to decode the interferometric data, the most effective
approach is a ``step-by-step'' strategy -- the complexity of the
models used to fit the data should increase gradually in
directions which are suggested by a thorough analysis of the simpler
model.
 
\begin{figure}
\begin{center}
\hspace{-2in}
\mbox{\epsfig{file=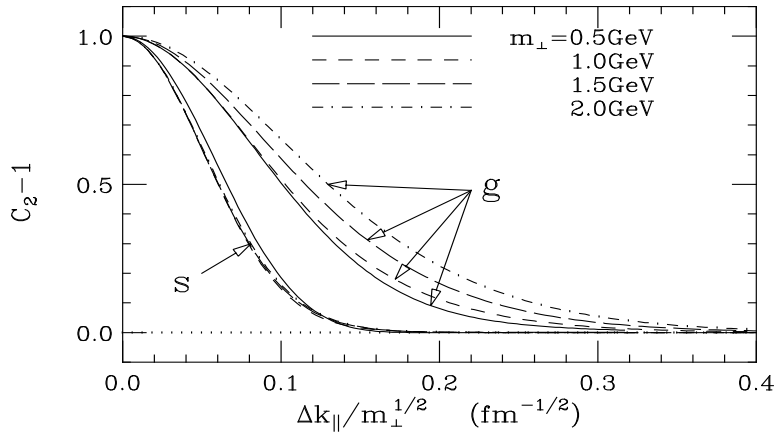,bbllx=185pt,bblly=321pt,bburx=465pt,bbury=687pt,angle=90,width=8cm,height=5.5cm}}
\vspace{-0.2in}
\end{center}
\caption{The correlation function $C_2$, as a function of
the rescaled longitudinal momentum difference.}\label{welke_fig:salve}
\end{figure}

It has been conjectured [\ref{welke:destil}] that kaons may escape
from the source earlier and over a longer period of time than pions.
Such a ``dynamical'' decoupling will influence the interferometric
data.  Suppose first that we have a sharp and common freeze--out of
all particles; taking $T_c=130~{\rm MeV}$ and $\tau=30~{\rm fm}$, we
obtain the curves marked ``s'' in Fig.~\ref{welke_fig:salve}. As
expected, we find that $\gamma \equiv a_{\scriptscriptstyle eff}
m_\perp^{1/2}$ is independent of $m_\perp$, and $\gamma_\pi=\gamma_K$.
Suppose on the other hand that the kaons decouple dynamically from the
pion fluid over some interval from $\tau_0$ to $\tau$; if
$\tau_0=7~{\rm fm}$ ($\tau T^3={\rm const}$ then implies $T_0=180~{\rm
  MeV}$), and assuming for simplicity an unweighted average of
emission function (\ref{welke_eq:E3.24}), $\langle
J(\tau)\rangle_\tau$, we obtain the curves marked ``g.'' Clearly,
$m_\perp$--scaling is violated for this type of emission. NA44 data
[\ref{welke:NA44}] gives no indication of such scaling violation,
suggesting that the kaon fluid remains coupled to the pions right
until their common (sharp) freeze--out. A systematic and parallel
study of pion and kaon interferometric source sizes as a function of
$m_\perp$ will provide a sensitive test for the temporal interval of
freeze-out in hadronic matter at RHIC.
\vspace{0.2in}

The dependence (\ref{welke_eq:E1}) may well apply at the SPS energies,
since most of the matter will cool and reach the freeze-out stage due
to the fast longitudinal expansion rather than the much slower
transverse expansion, but at the AGS it is well known that transverse
flow plays an important role. We illustrate the effects one might
encounter by way of two unphysical, but exactly calculable examples:

\noindent{\bf 2. Scale--invariant 2--d flow} and the 
explosion of a long filament: To find a possible signature of
transverse flow, consider the extreme case of a transversely expanding
filament of length $L\gg R_\perp$.  A convenient parameterization of
the coordinates is $x^{\mu} \;=\; (\tau\, \cosh\beta,\:
\tau\,\sinh\beta\,\cos\psi,\: \tau\,\sinh\beta\,\sin\psi,\: z)$, where
$\beta$ is the radial rapidity of the fluid cell. The temperature and
velocity field may be written as $\tau^2 T_c^3 = {\rm const}$, and
$u^{\mu} = (\cosh\beta,\: \sinh\beta\,\cos\psi,\\
\sinh\beta\,\sin\psi,\: 0)$, while
$d\Sigma^{\mu} =
u^{\mu}\,\tau^2\,\sinh\beta\: dz\,d\beta\, d\psi$.  The emission
function $J(k_1,k_2)$ may be calculated exactly for $m_{z1}=m_{z2}$,
where $m_z=\sqrt{p_z^2+m^2}$ is the longitudinal mass. We obtain
\begin{eqnarray}
  k^0\, {{dN_1} \over {d{\vec k}}} &=& J(k,k) \;=\; 2\pi L \tau^2 T_c
  \: {{\rm e}^{-m_z/T_c}\over (m_z/T_c)}\: \bigg [{m_z \over T_c}-1
  \bigg ]~ ~ ~,\label{welke_eq:E5.8} \\ 
  C_2(k_1,k_2)-1 &=& {\rm Re}\;{ {{\rm e}^{-2m_z
  (H-1)/T_c}\over H^6} \left[{ {m_zH/T_c +1} \over {m_z/T_c +1}
  }\right]^2 \left[1+{Q^2 \over 8 m_{z}^{2}} \right]^2 }~ ~ ~,
\label{welke_eq:E5.9}
\end{eqnarray}
\begin{equation}
{\rm where} ~ ~ ~ ~ ~
  Q^2=-(k_1-k_2)^2,~~~ H=\bigg [ 1+T_c^2\tau^2
  {Q^2\over m_z^2} -i{T_c\tau Q^2 \over 2m_{z}^{2}}\bigg ]^{1/2}~.
  \label{welke_eq:E5.6} 
\end{equation} 
Similar to the case of 1--d boost invariant flow, we obtain a plateau
for cylindrical boost invariant expansion, but now in the radial
rapidity distribution. The dependence of the spectral density on $m_z$
is noteworthy: the localization of the spectrum due to radial flow is
more pronounced for greater $m_z$. Particles with $m_z/T_c\gg 1$ spend
almost all their thermal energy for the longitudinal motion and thus
are strongly frozen into collective radial flow.  For this reason we
may expect the width of the correlator in the transverse direction to
be defined by $m_z$, rather than $m_{\bot}$.  This simple and
well--understood dependence leads us to conclude that $m_t$ scaling of
the transverse radius cannot be due to hydrodynamic expansion, and it
is even doubtful that it is at all possible.

In the case of small differences in radial particle
rapidity, the exact result (\ref{welke_eq:E5.9}) may be simplified to
obtain the same ``radii'' for the sideways and outward directions:
\begin{equation}
  R_{exp}=\tau\sqrt{T_c\over m_z},\;\;\;\; R_{cos}= {1\over
    T_c\tau}R_{exp}~ ~ ~,
\label{welke_eq:E5.12}
\end{equation}
\noindent where $R_{exp}$ dominates the shape of the 
correlator for $T_c\tau>1$, while $R_{cos}$ dominates for $T_c\tau<1$.
If the radial flow takes place against the background of a strong
longitudinal expansion, which is likely in a more realistic scenario
of the heavy ion collision, the distinctive $m_z$--dependence in the
data could be only washed out, but not replaced by an
$m_\bot$--dependence.

\noindent{\bf 3. Scale--invariant 3--d flow} and 
the explosion of a point--like source: For the case of spherical
expansion, $x^{\mu} = \tau\:(\cosh\beta,\:
\sinh\beta\,\sin\theta\,\cos\psi,\:
\sinh\beta\,\sin\theta\,\sin\psi,\: \sinh\beta\,\cos\theta)$, with
$\tau T_c= {\rm const}$, $u^{\mu}\;=\; x^{\mu} /\tau$, and
$d\Sigma^{\mu}\;=\; u^{\mu}\, \tau^3\, \sinh^2\beta \,\sin\theta \:
d\beta \, d\theta\, d\psi$.  Once again, the emission function
$J(k_1,k_2)$ may be evaluated exactly, and we obtain:
\begin{eqnarray}
  k^0\: {{dN_{1}} \over {d{\vec k}} } &=& J(k,k)\;=\; 4\pi \tau^3 T_c
  K_2({m\over T_c})~ ~, \label{welke_eq:E5.16} \\ 
    C_2(k_1,k_2)-1 &=& {\rm
    Re}{\left[ {K_2(H) \over H^2 K_2(m/T_c)}\right]^2 \left[1+{Q^2\over
        8 m^{2}}\right]^2}~,~ \mbox{{\rm where}} ~ H \equiv
  \sqrt{1+T_c^2\tau^2 {Q^2\over m^{2}} -i{T_c\tau Q^2 \over 2m^{2}}}~,
  \nonumber
\end{eqnarray}
and $K_2$ is the modified Bessel function.  We see that spherically
symmetric boost-invariant expansion washes out any inhomogeneity of
the local thermal spectrum. This is a distinctive feature of the
spherical model.  Approximating the correlator in the case of small
differences in radial particle rapidity, we obtain
\begin{equation}
  C_2(Q)-1 \;=\; {\rm e}^{- T_c\tau^2 Q_{out}^2/m}\:{1+Q^2/4m^2 \over
    1+5T_c^2\tau^2 Q^2/2m^2} \; \cos\bigg ({Q^2\over m^2}({5T_c\tau \over
    2} +{m\tau\over 2})\bigg )~ ~ ~.
\label{welke_eq:E5.18}
\end{equation}
All comments related to Eq.~(\ref{welke_eq:E5.12}) apply here, too.
Here, however, all three directions are physically equivalent and
the mass of the particle becomes the only parameter that may re-scale
the correlation radius.

\section{Conclusion}\label{welke_sec:conc}

We have emphasized some important physical aspects of the theory of
interferometry, and tried to isolate some controversial points.
Strictly speaking, interferometry does not permit the initial data to
be given in terms of semi--classical distributions, unless this
description is augmented by a clear indication of the length scale
that defines the quantum states of the particles.  Our conclusion is
that the traditional operator approach, based on the precise
definition of the particle states, provides a firm footing for the
calculation of the two--particle spectra.

Recent preliminary results by NA35 and NA44 seem to confirm the main
predictions of interferometry for a scale--invariant 1--d flow
scenario.  They provide a multidimensional test of the source, and the
fact that so many parameters coincide can hardly be accidental.  Thus,
we have strong evidence that at SPS energies a hydrodynamic regime
develops, and that freeze--out takes place during a short interval.
This collective behavior can be expected to occur at RHIC energies,
and thus it is highly desirable to continue developing the formalism.

We have extended previous calculations [\ref{welke:MS}] for a
longitudinally expanding system to the case of unequal transverse
momenta, and have shown that the HBT correlator may carry information
about the time of the kaon emission. We further considered
transversally expanding sources, and demonstrated how to change the
parameterization of the correlator.  Unfortunately, the attractive
feature that the spectrum is localized in the boost--invariant
solution for 1--d flow is absent in 2-- and 3--dimensional flows, even
if the radial motion is very strong.  This happens because the angular
dependence of the Cartesian velocity components is weak, and
practically means that one should perform all integrations exactly.
While even moderate radial flow does obscure the true transverse
source size, the spectrum is not sufficiently localized to allow one
to obtain a simple formula for the correlator (at least in terms of
standard variables). Numerical calculations are, of course, possible,
but then it becomes difficult to ``recognize'' the model.

We emphasize that while HBT for hydrodynamic sources is well
understood as a physical phenomenon, choosing an adequate model to fit
the data always poses difficulties. We impose the condition that the
model should allow one to recognize it via a qualitative analysis.
Only then can one hope to understand what parameters are responsible
for the behavior of the correlator, and find their values by fitting
the data.  Practically, this requirement means that we must begin with
an analytic expression for a solution of the relativistic hydrodynamic
equations. This guarantees the consistency between the shape of the
freeze--out surface and the velocity field. Unfortunately, analytic
solutions for the case of three--dimensional expansion are not yet
known. A reasonable analytic approximation will do, but to our
knowledge, a suitable expression has not yet been derived. An
approximate formula describing a realistic, expanding system at the
freeze--out stage is an important problem for boson interferometry at
RHIC. 

\noindent{\bf Acknowledgment:} This work was supported in part by the
U.S. Department of Energy under Contract No.~DE--FG02--94ER40831.

\section{References}

\begin{enumerate}
\item \label{welke:long} Based on: {A.~Makhlin,
    E.~Surdutovich, and G.~Welke}, Preprint WSU-NP-94-1 (1994).
\item \label{welke:review} See, for example: M.I.~Podgoretskii,
  Sov.~J.~Part.~Nucl.\, {\bf 20}, 266 (1989).
\item \label{welke:Mak72} A.~Makhlin, Sov.~Phys.~JETP\, {\bf 35}, 478
  (1972).
\item \label{welke:Scott} S.~Pratt, Phys.~Rev.~Lett.\, {\bf 53}, 1219
  (1984).\footnote{For an expanding hydrodynamical shell the 
  ``visible'' source size is smaller the larger the pair's total
  momentum. In nuclear collision models, the collision dynamics may
  result in a more complicated dependence: G.~M.~Welke, {\it et al.},
  in {\em Proc. of the
    10th Winter Workshop on Nuclear Dynamics}, January 15-22, 1994,
  Snowbird, Utah (World Scientific, Eds. J.~Harris, A.~Mingerey, and
  W. Bauer) p.~93.}
\item \label{welke:NA35}  The NA35 collaboration: R.~Morse {\it et
  al.}, Preprint LBL--36062 (1994).
\item \label{welke:NA44} The NA44 collaboration: H.~Beker
  {\it et al.}, Preprint CERN--PPE/94--119 (1994).
\item \label{welke:MS} A.~Makhlin, and Yu.~Sinykov,
  Sov.~J.~Nucl.~Phys.\, {\bf 46}, 354 (1987);\\ V.~Averchenkov,
  A.~Makhlin, and Yu.~Sinykov, Sov.~J.~Nucl.~Phys.\, {\bf 46}, 905
  (1987).
\item \label{welke:Phot}  
  A.~Makhlin, Sov.~J.~Nucl.~Phys.\, {\bf 49}, 238 (1989).
\item \label{welke:destil} J.~Rafelski, Phys. Reps.\, {\bf 331}
  (1982).
\end{enumerate}
\end{document}